\documentclass[sigconf,screen]{acmart}
\usepackage[utf8]{inputenc}
\usepackage{booktabs}
\usepackage[utf8]{inputenc}
\usepackage{lipsum}
\usepackage{graphicx}
\usepackage{subcaption}
\usepackage{array}
\usepackage{flushend}

\title[Espresso vs. EyeAutomate: Comparison of Two Generations of Android GUI Testing]{Fragility of Layout-Based and Visual GUI Test Scripts: An Assessment Study on a Hybrid Mobile Application}
\date{May 2019}

\begin{document}

\acmPrice{}
\acmDOI{10.1145/3340433.3342824}
\acmYear{2019}
\copyrightyear{2019}
\acmISBN{978-1-4503-6850-6/19/08}
\acmConference[A-TEST '19]{the 10th ACM SIGSOFT International Workshop on Automating TEST Case Design, Selection, and Evaluation}{August 26--27, 2019}{Tallinn, Estonia}

\author{Riccardo Coppola}
\affiliation{%
\institution{Politecnico di Torino}
\streetaddress{Corso Duca degli Abruzzi 24}
\city{Turin}
\state{Italy}
\postcode{10129}
}
\email{riccardo.coppola@polito.it} 
\author{Luca Ardito}
\affiliation{%
\institution{Politecnico di Torino}
\streetaddress{Corso Duca degli Abruzzi 24}
\city{Turin}
\state{Italy}
\postcode{10129}
}
\email{luca.ardito@polito.it} 
\author{Marco Torchiano}
\affiliation{%
\institution{Politecnico di Torino}
\streetaddress{Corso Duca degli Abruzzi 24}
\city{Turin}
\state{Italy}
\postcode{10129}
}
\email{marco.torchiano@polito.it}

\begin{abstract}
Context: Albeit different approaches exist for automated GUI testing of hybrid mobile applications, the practice appears to be not so commonly adopted by developers. A possible reason for such a low diffusion can be the fragility of the techniques, i.e. the frequent need for maintaining test cases when the GUI of the app is changed.

Goal: In this paper, we perform an assessment of the maintenance needed by test cases for a hybrid mobile app, and the related fragility causes.

Methods: We evaluated a small test suite with a Layout-based testing tool (Appium) and a Visual one (EyeAutomate) and observed the changes needed by tests during the co-evolution with the GUI of the app.

Results: We found that 20\% Layout-based test methods and 30\% Visual test methods had to be modified at least once, and that each release induced fragilities in 3-4\% of the test methods.

Conclusion: Fragility of GUI tests can induce relevant maintenance efforts in test suites of large applications. Several principal causes for fragilities have been identified for the tested hybrid application, and guidelines for developers are deduced from them.

\end{abstract}



\maketitle

\section{Introduction and Related Work}

Recent years have witnessed an enormous growth of mobile applications, both in capabilities offered and diffusions among end users. Mobile apps are generally divided in three main ways: \emph{Native} apps are developed using specific SDKs for the destination platform; \emph{Web-based} apps are typically developed with HTML5, XHTML Mobile Profile or JavaScript, and are engineered to be loaded in the browsers of mobile devices; finally, \emph{Hybrid} apps apps leverage both web and native technologies, using some native components to embed content that is created through web development technologies. 

Many different frameworks exist for developing hybrid applications: examples are Cordova \cite{bosnic2016development}, Xamarin, Flutter, React Native. In general, hybrid mobile development has the advantage of reducing the effort for developing cross-platform applications: native apps have to be developed and maintained for each OS separately, while with hybrid applications this applies only for few native components \cite{jabangwe2018software}. This advantage comes at the cost of slightly reduced performance \cite{dalmasso2013survey}, limited platform-specific features and, in some cases, reduced user's appreciation \cite{joorabchi2013real}.

Well-designed apps should guarantee high usability and should expose few defects to their end users. Hence, an in-depth testing of their GUIs (i.e., Graphical User Interfaces), through which most of the features are exposed, becomes fundamental. Specialized literature has however highlighted a scarce adoption of testing tools among all categories of mobile applications \cite{coppola2017scripted}\cite{7102609}. Among the main difficulties discouraging mobile testing, mobile apps are particularly prone to GUI testing Fragility, due to a rapid evolution of their GUI. A GUI test case used for regression testing is said to be fragile when failures are triggered not because of the injection of defects in the AUT, but for (even minor) changes in the GUI pictorial appearance or definition \cite{Coppola2019}.

Hybrid development exacerbates testing issues that are proper of mobile apps in general \cite{linares2017continuous} (especially the issue of Device Fragmentation, i.e. the number of different devices and/or operating systems on which apps have to be tested). Still, testing of hybrid mobile applications is discouraged because of the absence of full-fledged web testing frameworks specialized for hybrid apps, such as those available for native apps (e.g., Espresso for Android apps, and UI Automation for iOS ones) \cite{7226668}. A few tools, mainly adaptations of web application testing tools (e.g., Selenium) are available for testing hybrid applications.

The aim of this study is to perform an assessment of two different techniques for testing hybrid applications on a popular, large-sized and real application. In a previous work of ours, we already provided a comparison of the two techniques, in terms of the usability and quality of delivered test suites on a native application \cite{Ardito:2019:EVE:3319008.3319022}. For the purpose of this experiment, we developed test scripts for the first release of the application, and followed the release history maintaining the test cases in order to make them co-evolve with the production code, analyzing the adaptability of the considered approaches to the selected case study and evaluating the affordability, in terms of maintenance effort, of GUI testing for hybrid apps. 

The remainder of the present manuscript is organized as follows: section 2 summarizes the design of the study and provides details about the selected software objects; section 3 reports the results and gives answers to the defined Research Questions; section 4 reports the Threats to the validity of the current study; section 5 reports the findings of related works available in literature; section 5, finally, discusses the implications of the experiment and provides hints for future work.

\section{Study Design}

We report the design, goal, research questions, metrics and procedure adopted for the study following the guidelines by Wohlin et al. \cite{claes2000experimentation} on reporting empirical studies in Software Engineering.

The experiment can be described using the Goal Question Metric (GQM) paradigm \cite{caldiera1994goal}, as summarized in table \ref{tab:gqm}. The \emph{goal} of the experiment was to analyze the GUI testing process in the \emph{context} of hybrid mobile applications, to understand what kind of fragilities can be encountered when using two different techniques of testing techniques: \emph{2nd generation} or \emph{Layout-based}, and \emph{3rd generation} or \emph{Visual}. The results of the study are interpreted according to the perspectives of the \emph{developers} and \emph{testers} of Android apps, as well as \emph{researchers} aiming at mitigating the issues associated with the considered testing tools or techniques. The \emph{software object} of the experiment is a closed-source app detailed in a following paragraph.

\begin{table}
\small
    \centering
        \caption{GQM Template for the study}

    \begin{tabular}{@{}r@{~:~}l@{}}
    \toprule
        Object of Study & GUI Testing Approaches\\
        Purpose & Investigate needed maintenance in evolving test suites\\
        Focus & Fragility\\
        Context & Hybrid mobile applications\\
        Stakeholders & Developers, Testers, Researchers\\
        \bottomrule
    \end{tabular}
    \label{tab:gqm}
\end{table}

\subsection{Software Object}

\begin{figure*}[!t]
\centering
\begin{minipage}{0.23\textwidth}
  \centering
        \includegraphics[width=\textwidth]{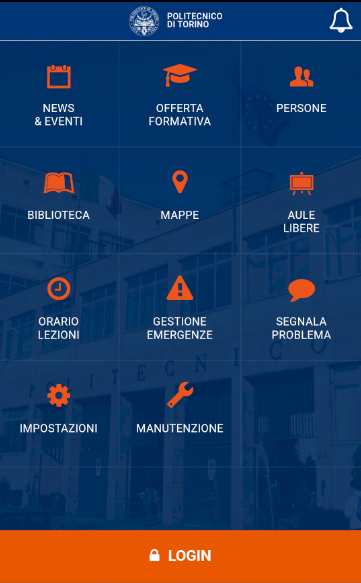}
        \label{fig:screens1}
        \subcaption{Home Menu}
\end{minipage}%
\hfill
\begin{minipage}{0.23\textwidth}
  \centering
        \includegraphics[width=\textwidth]{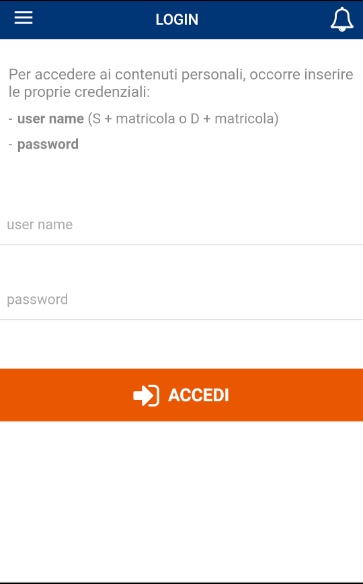}
                \label{fig:screens2}

        \subcaption{Login Form}
\end{minipage}%
\hfill
\begin{minipage}{0.23\textwidth}
  \centering
        \includegraphics[width=\textwidth]{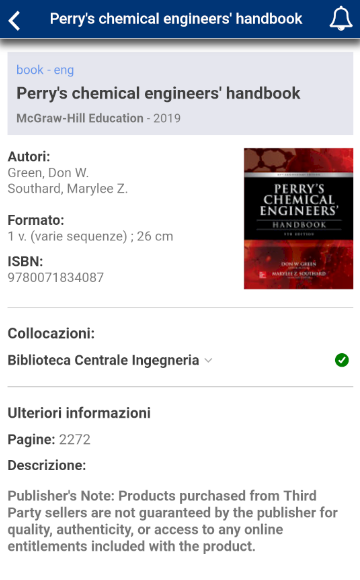}
                \label{fig:screens3}

        \subcaption{Library}
\end{minipage}
\hfill
\begin{minipage}{0.23\textwidth}
  \centering
        \includegraphics[width=\textwidth]{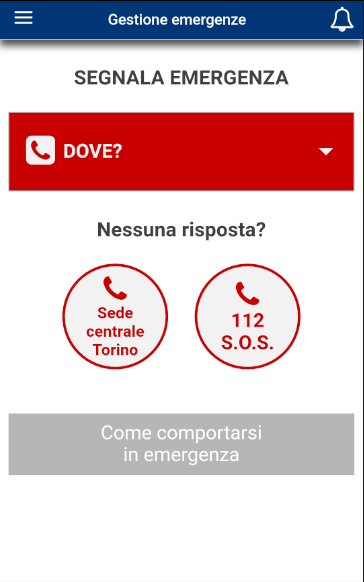}
                \label{fig:screens4}
        \subcaption{Emergency Signaling}
\end{minipage}
    \caption{Sample screens of PoliTO App}
    \label{fig:screens}
\end{figure*}


The study is based on the official mobile app of the Polytechnic University of Turin. The Italian language version of the app was used for the study: some sample screens are shown in figure \ref{fig:screens}. The app, named PoliTO App, is a hybrid application developed with Apache Cordova, and released on the three main app stores: Play Store \footnote{https://play.google.com/store/apps/details?id=it.polito.politoapp}, App Store \footnote{https://itunes.apple.com/it/app/polito-app/id1087287751}, and Microsoft Store \footnote{https://www.microsoft.com/it-it/p/polito-app/9nblggh4n7f2}. The app allows both students and professor/researchers to login (see figure 1.b) and access a set of specific features: news, maps, lectures timetable, issue reporting (see figure 1.d), info about transportation, info about degree courses, info about people working in the University, info about books available in library (see figure 1.c), contacts, for both types of users; management of exam calls, list of work assignments, for professor and researcher users only; info about the career and the possibility of booking exam calls, for student users only. All the features are reachable from a grid-based home activity.

The app is popular among students enrolled to the University and has a user base of about 40 thousand users (around 80\% of which having Android devices). The application history is made of 26 releases, with release 1.7.2 -- the latest available when the study was performed -- published on February 21st, 2019. The app has gathered very good feedbacks from the users, with a rating of 4.4/5 stars on Google Play (average of 1,005 reviews) and of 4.5/5 stars on the App Store (average of 65 reviews).

\subsection{Testing Tools}

Two testing tools were adopted to develop test suites for the described software object.

Appium\footnote{http://appium.io/} \cite{shah2014software} is an open-source tool that can be used for testing both Android and iOS applications. According to a classification of GUI testing tools provided by Alegroth et al. \cite{alegroth2015conceptualization}, Appium belongs to the category of \emph{2nd generation} or \emph{Layout-based} testing tools, meaning that it uses layout properties of the on-screen widgets as locators (to identify elements to interact with) or oracles (to verify the correct execution of the tested usage scenario). It leverages UIAutomator and UIAutomation for accessing native components, and Selendroid, WebDriver and ChromeDriver for interacting with generated web components of Hybrid apps. Appium test scripts can be developed in different languages (e.g., Java, JavaScript, Python or Ruby) and can be obtained through manual scripting or basic functionalities of capture \& replay (through the Appium Recorder plug-in). Appium supports image recognition to generate test cases leveraging locators and oracles of both 2nd and 3rd generation.

EyeAutomate\footnote{http://eyeautomate.com} \cite{alegroth2018continuous} is an open-source testing library that belongs to the category of \emph{3rd generation} or \emph{Visual} testing tools. It applies visual recognition on actual screen captures of the application to locate elements on which to interact and to check oracles of the proper behaviour of the application. The tool allows to automate any application provided with a GUI, and hence it can be applied to mobile apps if they are emulated on the screen of a desktop pc. The EyeAutomate API allows the development of test methods in Java and the integration with test execution environment like JUnit. Additionally, tests can be created through Capture \& Replay, using a module embedded in the dedicated EyeStudio IDE to collect screen captures; EyeStudio allows to save test scripts in plain ".txt" files to be reproduced.

\subsection{Research Questions and Metrics}

To achieve the proposed goal of the study, we formulated the following research questions:

\begin{description}
\item[RQ1:] Were test suites written with 2nd and 3rd generation tools able to spot defects in the application?
\end{description}

To answer this question, we analyzed failing tests to understand whether the failures were caused by new defects of the evolving application.

\begin{description}
\item[RQ2:] What is the amount of maintenance effort needed by the Layout-based and Visual test suites?
\end{description}

To quantify the maintenance effort, we applied some of the change metrics that we defined for mobile tests in an earlier work \cite{8477182}. In particular, we measured the following metrics:

\begin{itemize}
    \item \emph{$MMR_i$ (Modified Methods Ratio)} is defined for release $i$ as $MM_{i} / TM_{i-1}$, where $MM_{i}$ is the number of modified test methods in the transition between releases $i-1$ and $i$, and $TM_{i-1}$ is the total number of test methods in release $i-1$. This metric gives information about the ratio of test methods that require manual intervention during the evolution of a tested app.
    \item \emph{$TMV_j$ (Test Method Volatility)} defined for test method $j$ as $Mods_j / Lifespan_j$ where $Mods_j$ is the amount of releases in which the method $j$ is modified, and $Lifespan_j$ is the number of releases of the app package that feature the test method $j$.
        \item \emph{$TSV$ (Test Suite Volatility)} defined as the ratio between the number of methods that are modified at least once in their lifespan (hence, having TMV > 0) and the total number of methods;
        
        \item \emph{$MRR$ (Modified Releases Ratio)} defined as the ratio between the number of releases in which at least a test class associated to a given GUI automation framework has been modified, and the total amount of releases.
\end{itemize}

Average $MMR$ value was eventually computed on the full set of releases of the app. Average $TMV$ value was eventually computed on the full set of test methods.

\begin{table}
\small
    \centering
        \caption{Metrics defined to answer RQ2}

    \begin{tabular}{cc}
    \toprule
    Metric Name & Meaning\\
    \midrule
    MMR & Modified Methods Ratio\\
    TMV & Test Method Volatility\\
    TSV & Test Suite Volatility\\
    \bottomrule
    \end{tabular}
    \label{tab:metrics-acronyms}
\end{table}

Table \ref{tab:metrics-acronyms} summarizes the adopted metrics and their meaning.

\begin{description}
\item[RQ3:] What kind of fragilities are encountered during the evolution of a Hybrid Mobile app?
\end{description}

To answer this question, each time a test failed because of changes in the AUT's definition or its GUI, we analyzed the reasons and the needed modifications. Such causes were then assigned to one of the categories of the taxonomy of modification causes that we defined in an earlier work and took as a reference \cite{8411747}.

Number and frequency of occurrences of each (macro)category were then measured to provide an answer to RQ1.

To compare the failures of test cases caused by fragilities with legitimate failures caused by defects in the app, we also considered the situations in which the test cases failed due to triggered bugs.

\subsection{Experimental Procedure}

{

\begin{table}[t]

\small
    \centering
        \caption{Description of Usage Scenarios}
    \begin{tabular}{ll}
         \toprule
         \textbf{Name} & \textbf{Description} \\
         \midrule
         T1 & Search People - Wrong Input\\
         T2 & Search People - Correct Input\\
         T3 & Check Contacts\\
         T4 & Base Issues Signaling\\
         T5 & Library - Search by Author\\
         T6 & Library - Search by Title and Year\\
         T7 & Check Notifications\\
         T8 & Student Login\\
         T9 & Student - Check Courses\\
         T10 & Student - Check Exams Taken\\
         T11 & Logout\\
         T12 & Professor Login\\
         T13 & Professor - Check Work Assignments\\
         T14 & Professor - Check Exam Calls\\
         T15 & Professor - Extended Issues Signaling\\
         \bottomrule
    \end{tabular}
    \label{tab:scenarios}
\end{table}

}

The main usage scenarios of the app were exercised with fifteen different test cases, listed in table \ref{tab:scenarios}. 

Test scripts were not defined for usage scenarios producing output that rapidly changes over time, based on time/date or device location (e.g., News Feed, Lectures Timetable, Public Transportation, Available Rooms).

Several usage scenarios required the Login (as student or professor user) as a precondition. Those test cases did not include the login procedure, which was performed only by specific test cases that performed the login only. This introduced an order relationship between test cases: T9, T10, and T11 had to be executed after T8; T13, T14, T15 had to be executed after T12.

The layout-based test suite with Appium was created as a single Java class, with each usage scenario covered by a dedicated test method. ChromeDriver was used to automate all the interactions with the WebView of which the whole screen of the app is composed. XPath expressions, with checks based on different parameters, were used to identify the elements to interact in the usage scenarios. Few elements were provided with a unique id; hence, the preferential locator used was the class name of the widgets. In absence of unique class names, the text contained inside the widget (or the textual placeholder which is shown before any text input in a text box) was used as a locator. When text or placeholders were used as locators, no condition was set on the type of class in the XPath expression. Multiple locators (typically, combinations of class and text locators) were used when the usage of text and placeholder locators only led to ambiguities in the identification of screen elements. Due to the way the app was built, it was not possible to identify a unique identifier to the \emph{Navigate back} button (in the top-left corner of the screen). We performed the related operation by directly sending the \emph{BACK} keystroke event, which in a connected Android device emulates a tap on the back button.

\begin{figure}
    \centering
    \includegraphics[width=0.35\columnwidth]{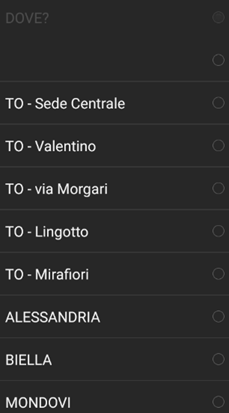}
    \caption{Defect detected by Visual test case T3}
    \label{fig:bugvisualt3}
\end{figure}

The visual test suite with EyeAutomate API was created as well as a single Java class, with methods corresponding to the different usage scenarios. The screen captures were taken by using the image capture tool embedded in the EyeStudio IDE. The recognition mode of the tool was set to \emph{FAST}, meaning that the image recognition was based on color changes and not on contrast changes (the default \emph{TOLERANT} mode). The image recognition library was incapable of identifying an exact image capture of the \emph{Navigate back} button, possibly because of its small size and relative scarce amount of graphic information. To cope with this issue, an extended capture was taken to consider the whole upper navigation bar, and to click on its leftmost part. 

All test cases developed with both tools included sleep instructions between each pair of commands; we inserted a sleep of 2 seconds after each operation that caused a navigation to another screen of the app, and a sleep of 0.5 seconds after each operation that caused changes in the visual hierarchy of the current screen without triggering the transition to another one.

All tests were performed on an emulated Nexus 5X API 25 (Android 7.1.1), screen size 1080x1920:420dpi, with enabled device frame and enabled keyboard inputs. Tests were run with Windows 10 as host system, with screen resolution of 1920x1080.

The test scripts were first created for release 1.1.0 of the application, and generated after a correct execution of the related usage scenario. The test suite was then applied to the subsequent release of the application, to check whether test scripts were failing due to fragilities or regression in the AUT. This procedure was iterated on all the 25 releases of the application. For each release, the $MMR$ metric defined in section 2.3 was computed, and the encountered fragilities or bugs were classified and documented. Average $TMV$ and $TSV$ values were computed at the end of the iteration over the app releases.

\section{Results and Discussion}

This section reports the detected defects and fragilities during the execution of the test cases on all the available releases of the considered AUT, and the computed metrics for both the Layout-based and Visual test suites.

\subsection{Discovered Defects}

According to a distinction that is often adopted in literature, test cases can be defined as Passing, Failing or Flaky. A test is considered \emph{Failing} when all its executions end with a failure. A test is considered \emph{Flaky} when some of its executions end with a failure but not all~\cite{luo2014empirical}. Flakiness can be due to non-deterministic errors triggered during a test case execution, or to other conditions (e.g., availability of the AUT's backend, memory usage on the device, speed of the network connection) that may cause test case failures. We executed failing test cases multiple times (up to ten times) to discriminate between failing and flaky ones.

Several test cases were flaky because of non-deterministic behaviours of the AUT. Starting from release 1.5.2, the Layout-based version of test case 3 experienced several failures because the contacts were not properly loaded on-screen. 

Both Layout-based and Visual version of the test case 15 experienced failures in release 1.4.8, because of a needed swipe operations that happened to be not performed properly on screen. The issue ceased to cause failures in test cases from release 1.5.4 onwards, in which the layout of the home screen was re-arranged so that the swipe was no longer necessary on the used virtual device.

A single visualization defect was spotted by the Visual test case T3 when applied to release 1.6.4 of the AUT (see figure \ref{fig:bugvisualt3}): an empty element in a drop-down menu. The defect is possibly due to an initialization value of a List that is not removed after its population. The defect is not present in release 1.6.2, when the dropdown menu is first introduced in the Contacts feature of the app. The defect cannot be detected by the Layout-based test suite, which uses as oracles and locators the textual content of the dropdown menu voices. Instead, the related visual test case uses the screen capture of the entire dropdown menu as an oracle, and hence detects the missing element. At release 1.7.2 the defect has not been corrected by the developers of the AUT.

\vspace{1ex}
\noindent\fbox{%
    \parbox{0.97\columnwidth}{%
        \textbf{Answer to RQ1}: Several non-deterministic test failures were found by executing the developed test suites, related to issues in downloading information for a usage scenario and in performing a scroll operation on the home page of the app. The Visual test suite was able to detect a visualization bug in the Contacts section of the application, that was not fixed by the developers.
    }%
}\\[2ex]

\subsection{Measured maintenance effort}



\begin{figure}
    \centering
    \includegraphics[width=\columnwidth]{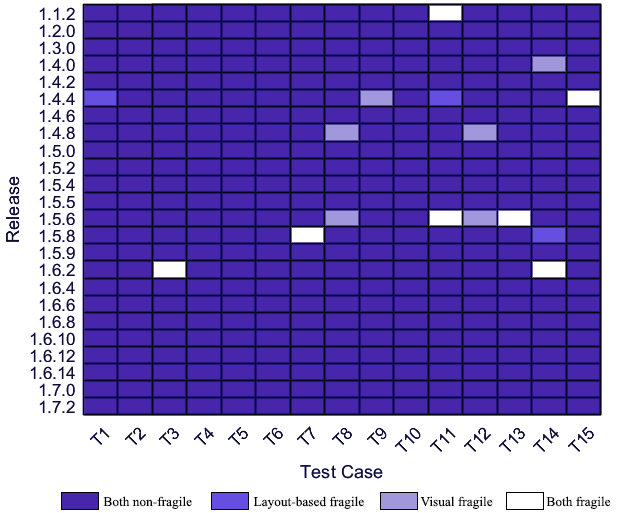}
    \caption{Non-fragile and fragile Layout-based and visual tests, during the co-evolution with the AUT}
    \label{fig:fragiletests_boh}
\end{figure}

\begin{table}
\small
    \centering
        \caption{Metrics for developed test suites}

    \begin{tabular}{crrrr}
        \toprule
        Test suite & $\overline{MMR}$ & $\overline{TMV}$ & TSV & MRR\\
        \midrule
        Layout-based & 2.8\% & 2.8\% & 47\% & 20.9\%\\
        Visual & 3.9\% & 3.6\% & 60\% & 29.2\% \\
        \bottomrule
    \end{tabular}
    \label{tab:metrics}
\end{table}

The colormap in figure \ref{fig:fragiletests_boh} reports the outcomes of the executions of the test cases during the evolution of the AUT, for the Layout-based and Visual test suites. The graph does not include release 1.1.0, for which the test cases were first defined and hence were all executable. 

Table \ref{tab:metrics} reports the average Modified Methods Ratio for all releases, the average Test Method Volatility, Test Suite Volatility and Modified Releases Ratio for both the developed Layout-based and Visual test suites.

For what concerns the Layout-based test suite, 7 out of 15 test cases (TSV = 47\%) were fragile at least one time during the lifespan of the AUT, with T11 being the most fragile with three required changes during the 25 considered releases. The average measured Test Method Volatility (TMV) was of 2.8\%. 5 out of 24 releases had at least one fragile test case, and hence a Modified Releases Ratio of 21\% could be measured.

Regarding the Visual test suite, 9 out of 15 test cases (TSV = 60\%) were fragile at least one time during the lifespan of the AUT, with T11 and T12 being fragile the most since they required maintenance two times each. The average measured Test Method Volatility (TMV) was of 3.6\%. 7 out of the 24 releases had at least one fragile test case, and hence a Modified Releases Ratio of around 29\% could be measured.

For both test suites, no maintenance was needed from release 1.64 up to the latest considered release (1.7.2). An inspection of the AUT and the related changelog confirmed that no significant interventions were performed in the features tested by the 15 considered test cases in those releases, with a focus on added features (e.g., a new Settings option from the main menu) instead.

\vspace{1ex}
\noindent\fbox{%
    \parbox{0.97\columnwidth}{%
        \textbf{Answer to RQ2}: Visual test cases needed a slightly higher amount of maintenance during their co-evolution with the release of the AUT. 20-30\% of releases needed maintenance effort due to fragilities in any of the test cases, and 47-60\% of developed test cases had to be maintained at least once during the lifespan of the app.
    }%
}\\[2ex]

\subsection{Fragility Causes}

\begin{table}
    \centering
    \small
            \caption{Fragility cause occurrences}

    \begin{tabular}{lrr}
    \toprule
    Fragility Cause & Layout-based & Visual\\
    \midrule
        Text Change & 5 & 3 \\
        Application Behaviour Change & 2 & 2\\
        \midrule
        Widget Substitution & 2 & 0\\
        Widget Removal & 1 & 0\\
        \midrule
        Graphic Change & 0 & 5\\
        Widget Arrangement & 0 & 2\\
        Widget Addition & 0 & 1\\
        \bottomrule
    \end{tabular}
    \label{tab:fragility_reasons}
\end{table}

%

%

Table \ref{tab:fragility_reasons} reports the causes of the encountered fragilities during the execution of Layout-based and Visual test cases. The causes for maintenance are taken from the taxonomy defined in our previous works \cite{8411747}.

\emph{Text Change} fragility happens when test cases are invalidated by changes in the textual content of elements of the GUI. Text change was the  most frequent fragility cause for the Layout-based test suite. This issue was related to the nature of the AUT, where the elements of the user interface were not provided with a unique identifier, and hence many of the locators were based only on the text contained by the widgets. Text-based locators are very prone to fragilities, because changes in the text of GUIs of mobile apps are likely to be frequent during the evolution of the project. The number of fragilities induced by Text Change was slightly lower for Visual test cases. This was mainly due to the robustness of the adopted image recognition algorithm to minor changes in the textual content of the screen captures used as locators, whereas a single different character is sufficient to invalidate a Layout-based locator based on String comparison.

\begin{figure}
\centering

\begin{minipage}{0.45\columnwidth}
  \centering
        \includegraphics[width=\textwidth]{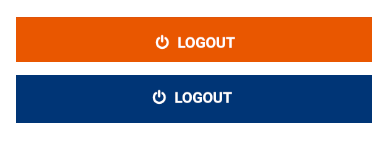}
        \subcaption{Visual TC11, rel. 1.1.2}
        \label{fig:widgetarrangement1}
\end{minipage}%
\hfill
\begin{minipage}{0.45\columnwidth}
  \centering
        \includegraphics[width=\textwidth]{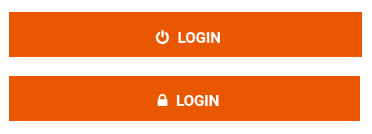}
        \subcaption{Visual TC8/12, rel. 1.5.6}
        \label{fig:widgetarrangement2}
\end{minipage}%

\caption{Examples of Pure Graphic Fragilities}

\label{fig:puregraphic}
\end{figure}

\emph{Graphic Change} fragilities are caused when the drawable associated to a widget changes between two consecutive release of the same application. This type of change in the AUT is as expected the primary cause of fragility for the Visual test suite, and does not cause any fragility in the Layout-based test suite. Examples of Graphic Change fragilities, related to the Login and Logout buttons, are reported in figure \ref{fig:puregraphic}.

\emph{Widget substitution} fragilities are caused when the class or type of the interacted widgets change between two consecutive release of the same application. The presence of this type of fragilities in the Layout-based test suite was again due to missing unique IDs in the AUT's GUI definition. Many locators, in fact, were based on the name of the class of the interacted widgets, that was subject to change in different releases. As predictable, the Layout-based test suite did not suffer from this type of fragility, since the image recognition of the widgets is completely agnostic of their definition.

\emph{Application Behaviour Change} fragilities are caused by changes in the expected behaviour of the app in a given usage scenario. Two usage scenarios, T3 and T13, changed during the evolution of the AUT. The respective test cases had to be changed accordingly -- with an adaptation of the locators as well as the oracles -- for both Layout-based and Visual test suites.

\begin{figure}
\centering

\begin{minipage}{0.45\columnwidth}
  \centering
        \includegraphics[width=\textwidth]{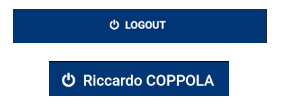}
        \subcaption{Visual TC11, rel. 1.5.6}
        \label{fig:widgetarrangement1}
\end{minipage}%
\hfill
\begin{minipage}{0.45\columnwidth}
  \centering
        \includegraphics[width=\textwidth]{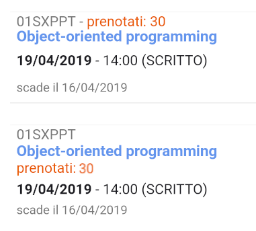}
        \subcaption{Visual TC13, rel. 1.4.0}
        \label{fig:widgetarrangement2}
\end{minipage}%

\caption{Examples of Widget Arrangement Fragilities}

\label{fig:widgetarrangement}
\end{figure}

\emph{Widget Arrangement Change} fragilities happen when the relative position on screen of the widgets to interact are changed between two different releases of the app. This fragility is experienced in two different test cases, which are reported as examples in figure \ref{fig:widgetarrangement}. In release 1.5.6 (figure \ref{fig:widgetarrangement1}), the Logout button is moved to the upper part of the screen, beside the name of the logged user. This change invalidates the screen capture used as a locator for the related Visual test script, while the Layout-based one (which uses the class of the logout button as a locator) does not need maintenance. In release 1.4.0 (\ref{fig:widgetarrangement2}) the arrangement of the widgets in a layout is changed. The related Visual test case, which uses as oracle the apperance of the layout as a whole, needs the oracle to be updated. On the other hand, the Layout-based test, which uses the unchanged text content as oracle, does not fail.

\emph{Widget Addition / Widget Removal} fragilities are caused by the addition or removal of elements in the GUI of the AUT. Such changes in the widget hierarchy may impact the assertions used in test cases (e.g., an assertion for the content of a no longer present widget) or visual oracles if the newly added widgets change the appearance of a whole layout used with that purpose.

\vspace{1ex}
\noindent\fbox{%
    \parbox{0.97\columnwidth}{%
        \textbf{Answer to RQ3}: Seven different reasons caused the need for maintenance in Layout-based and Visual test suites during the co-evolution with the application. The most frequent fragility cause for the developed Layout-based test cases was the modification of textual content of widgets, whereas the most frequent one for Visual test cases was the change of the drawables used for individual widgets of the GUI.
    }%
}\\[2ex]

\section{Threats to Validity}

\emph{Threats to Internal Validity}. The results of the study may be influenced by the selection of the usage scenarios that were covered by the developed test suite, and by the way test cases were developed (e.g., for the selected locators and oracles). The ''precision'' setting for the EyeAutomate library can also influence the results of the analysis. The number of test cases was also relatively small (if compared to industrial test suites) and based only on released versions of the application, hence possible defects promptly removed by the developers may have not been present in the inspected releases.

Researcher bias is another possible threat for comparative studies. The authors are not involved in the development of neither the considered app or the two considered testing tools, and have no reasons to favour any particular approach neither are inclined to demonstrate any specific result.

\emph{Threats to External Validity}. The results of this work are applicable only to the considered testing tools -- i.e. Appium and EyeAutomate -- and it is hence unsure if they can be generalized to other testing tools belonging to the same generations of GUI testing, and/or generating test scripts using the same approaches. The two tools however are quite widespread in literature and have several similarities to other common testing tools, like Selenium \cite{bruns2009web} for the Layout-based generation, and Sikuli \cite{yeh2009sikuli} for the Visual generation. It is not assured that the considered application, developed in Apache Cordova, can be deemed representative of all web-based hybrid mobile applications. Finally, the test cases were executed on a single Android Virtual Device: it is not sure whether the outcomes of the execution would have been the same if other devices were used.

\section{Conclusion and Future Work}


We performed an assessment study of two different testing techniques and tools on a widely used mobile hybrid application, developed with Cordova. On average, 2.8\% of Layout-based and 3.9\% of Visual test methods needed maintenance due to fragility in each new release of the application; around 20\% and 30\% of the test methods, respectively of Layout-based and Visual test suites, had to be modified at least once during the evolution of the app. Such measures are in line with the results of a previous mining study that we performed for Native Android applications released on GitHub, and hence confirm that Hybrid applications are subject to a comparable degree of fragility to that of Native Android apps. 

For what concerns the causes of fragility, we found that text changes were the principal fragility for the Layout-based test suite. This kind of fragility can be easily circumvented by adopting unique IDs for all the widgets in the GUI hierarchy of the AUT, to avoid relying on very volatile locators as the textual content of the widgets.

In addition to several cases of test flakiness, the developed Visual test suite was capable of detecting a visualization bug in the tested app, that was not fixed yet by the developers when the experiment was conducted. The test case was not detected by the analogous Layout-based test. This result suggests that a combined usage of Layout-based and Visual test suites should be considered a best practice for mobile apps, to be able to find defects in both the pictorial appearance of the GUI (that can be easily found by Visual test cases) and in the behaviour and relationships between specific widgets (that can be found by Layout-based test cases).

As our future work, we are aiming to implement and validate a combined, translation-based tool for Visual testing of mobile apps, in order to leverage the benefits of the two testing approaches and obtain enhanced defect detection. Since the fragility causes for Layout-based and Visual test cases are many times complementary, a translational approach can also allow to repair or generate fragile test cases using still working test cases of the other generation, hence reducing the maintenance effort. We plan to provide support for the translation to EyeAutomate and Sikuli scripts, starting from Espresso test cases for native Android apps \cite{ardito2018towards}, and from Appium for web-based hybrid apps. The translation-based approach has been already applied in the field of Web-Application testing, translating Selenium test scripts to Sikuli \cite{leotta2018pesto}.

We also plan to develop automated tools to automatically repair locators or oracles used by Layout-based and Visual tests, with an automated static analysis on the application package. Automated repairs of locators (in addition to automated refactoring of test code when test classes are developed in the same integrated development environment of the AUT) still appear missing in literature \cite{linares2017continuous}, and would constitute a valuable aid in reducing the amount of maintenance needed by test cases.


\bibliographystyle{ACM-Reference-Format}
\bibliography{sample-bibliography}

\end{document}